\documentclass[10pt,twocolumn,letterpaper]{article}

\usepackage{cvpr}
\usepackage{times}
\usepackage{epsfig}
\usepackage{graphicx}
\usepackage{amsmath}
\usepackage{amssymb}

\usepackage{booktabs}
\usepackage{latexsym}
\usepackage{subfigure}
\usepackage{amsmath,bm}
\usepackage[ruled]{algorithm2e}
\usepackage{setspace}
\usepackage{multirow}

\newcommand{\uiName}{dualFace}
\newcommand{\vcsize}{0.5cm}
\newcommand{\hzy}[1]{\textcolor{black}{#1}}
\newcommand{\hzyv}[1]{\textcolor{black}{#1}}

\newcommand{\argmax}{\mathop{\rm arg~max}\limits}
\newcommand{\argmin}{\mathop{\rm arg~min}\limits}

\newcommand\userid{1}

\usepackage[breaklinks=true,bookmarks=false]{hyperref}

\cvprfinalcopy 

\def\cvprPaperID{****} 

\begin{document}

\title{dualFace:Two-Stage Drawing Guidance for Freehand Portrait Sketching}

\author{Zhengyu Huang$^1$, Yichen Peng$^1$, Tomohiro Hibino$^1$, Chunqi Zhao$^2$,\\
 Haoran Xie$^1$, Tsukasa Fukusato$^2$, and Kazunori Miyata$^1$\\
$^1$Japan Advanced Institute of Science and Technology\\
$^2$The University of Tokyo}

\maketitle

\begin{abstract}
  Special skills are required in portrait painting, such as imagining geometric structures and facial parts' detail of final portrait designs, making it a difficult task for users, especially novices without prior artistic training, to draw freehand portraits with high-quality details. 
In this paper, we propose dualFace, a portrait drawing interface to assist users with different levels of drawing skills to complete recognizable and authentic face sketches. 
dualFace consists of two-stage drawing assistance to provide global and local visual guidance: global guidance, which helps users draw contour lines of portraits (i.e., geometric structure), and local guidance, which helps users draws details of facial parts (which conform to user-drawn contour lines), inspired by traditional artist workflows in portrait drawing. 
In the stage of global guidance, the user draws several contour lines, and dualFace then searches several relevant images from an internal database and displays the suggested face contour lines over the background of the canvas. In the stage of local guidance, we synthesize detailed portrait images with a deep generative model from user-drawn contour lines, but use the synthesized results as detailed drawing guidance. 
We conducted a user study to verify the effectiveness of dualFace, and we confirmed that dualFace significantly helps achieve a detailed portrait sketch.  \footnote{This is the author version of manuscript which has been accepted in Journal of Computational Visual Media. Code is available at https://github.com/shasph/dualFace.}

\end{abstract}

\section{Introduction}
\label{sec:intro}
Portrait painting is one important art genre to represent a specific human 
from the real world or one's imagination. Some artists, together with their famous portrait drawings, have been widely adored for hundreds of years (e.g., \textit{Mona Lisa} and \textit{Girl with a Pearl Earring}). 
However, drawing portraits is cumbersome and requires special skills and capabilities (for example, spatial imagination and essential drawing skills), which are inaccessible to novices without prior artistic training. Therefore, the present paper 
aims to establish a user-friendly framework to support the process of drawing freehand portrait.

Several systems have been proposed for supporting portrait drawings in guidance-based method. For example, Portraitsketch~\cite{XieHLW14} proposes a framework to display an artistic rendering sketch using tracing. However, the user must prepare a reference image in advance, which can be time consuming. Shadowdraw~\cite{lee2011shadowdraw} and Sketchhelper~\cite{choi2019sketchhelper} incorporate image retrieval methods with tracing tools to dynamically search the relevant images from a database instead of manual selection and enable users to understand geometric structures of target designs (e.g., facial parts' locations and proportions). Although approaches mentioned above can help users copy existing drawings, it is still difficult to explore ``new'' portrait designs. 
In addition, these systems are unsuitable for drawing the details of portraits (e.g., facial parts' details) because they simply blend a set of relevant images. That is, the details of each image might be lost. Conversely, to explore new drawing designs to explore new drawing designs (detailed drawings), Ghosh et al.~\cite{ghosh2019interactive} and Zhu et al.~\cite{zhu2016generative} employ deep learning methods, especially with generative adversarial networks (GAN), to generate possible images with given color or edge constraints. However, the resulting image quality is still determined by the user's drawing skill, such as locating facial parts, so it remains difficult for novices to design high-quality portrait drawings.

\begin{figure*}[t]
\centering
\includegraphics[width=0.85\textwidth]{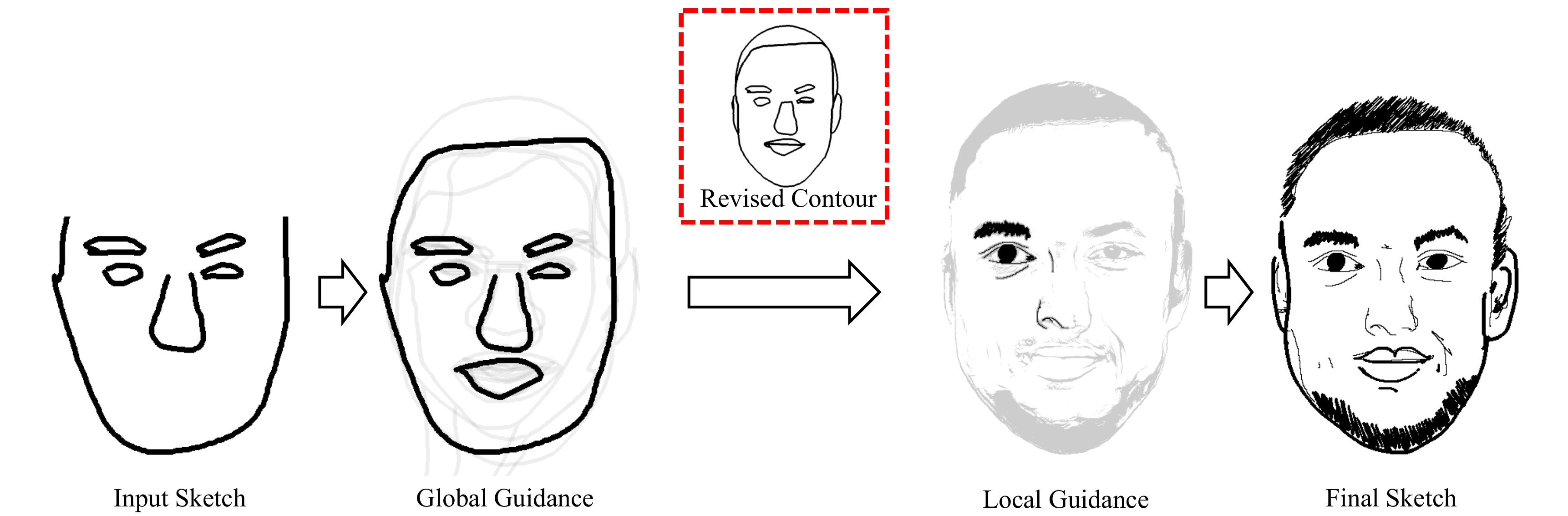}
\caption{The proposed portrait drawing interface provides both global and local guidance from the input of the user sketch. The revised contour sketch in the back end is from the merged mask generated by our conversion algorithm according to the input sketch, which is the reference for local guidance generation.}
\label{fig:concept}
\end{figure*}

To address the problems above, we referred to the conventional portrait drawing procedures~\cite{bradley2003draw}. According to the conventional procedures, it is essential for novices to adopt two types of guidance; (i) global guidance, which helps users locate facial parts (geometric structures) with correct proportions and (ii) local guidance, which helps users design facial details (e.g., eye and nose). Nonetheless, previous researches do not argue how to guide users to draw both global and local features of portraits, to our knowledge. 
Thus, we first consider a method to automatically generate two types of visual guidance, called global and local guidance, from user drawings (see Figure~\ref{fig:concept}). 
In case of global guidance, as with Shadowdraw~\cite{lee2011shadowdraw} and Sketchhelper~\cite{choi2019sketchhelper} mechanism, when the user draws contour lines on canvas, the system dynamically searches relevant images from a database and generates a blended image. 
In case of local guidance, the system generates detailed facial portraits from the user-drawn contour lines by using a GAN-based system and displays one of them. 
Second, we implement a graphical user interface (GUI), called \uiName, that incorporates the above visual guidance and is able to switch between the two stages freely.

Our principal contributions are summarized as follows.
\begin{itemize}
\item A two-stage guidance system that helps users design portrait drawings with data-driven global guidance and GAN-based local guidance. 
\item An optimization method to automatically generate detailed facial portraits with semantic constraints from user-drawn strokes. By using the generated portraits as drawing guidance, the user can explore the desired details without prior artistic training. 
\item A user study to demonstrate the benefits of our proposed system.
\end{itemize}

\begin{figure*}[t]
\centering
\includegraphics[width=0.85\textwidth]{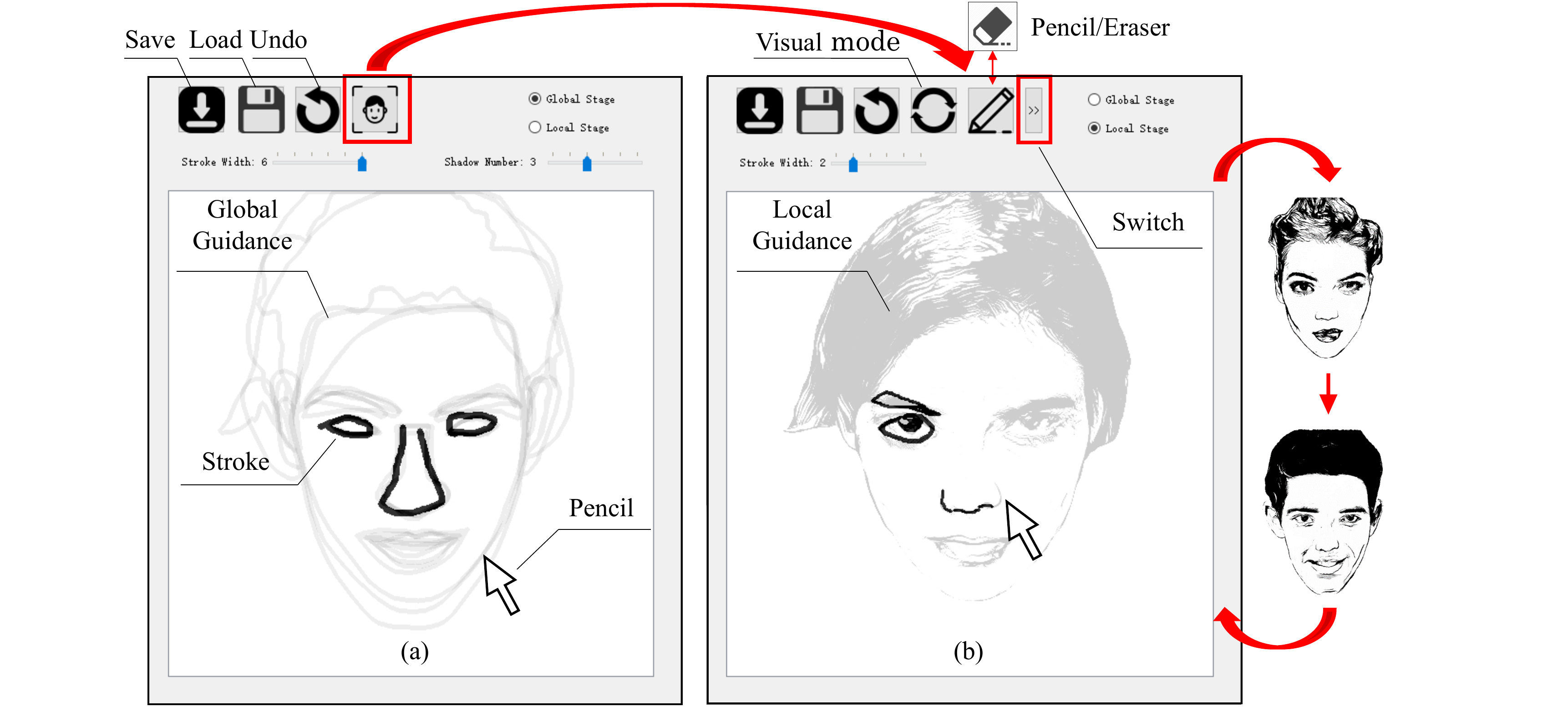}
\caption{Our two-stage guidance system. Given a user's intermediate drawing at runtime, the system generates (a)~global guidance generated by blending relevant images from the database and (b)~local guidance (i.e., realistic facial portraits) generated by a generative model-based method.}
\label{fig:ui}
\end{figure*}

\section{Related Work}
\subsection{Sketch-based Applications}
Sketch is a high-level abstract visual representation without lots of visual details. By analyzing the intention behind users' freehand sketch,
the sketch-based interaction allows users intuitive access to various applications such as image retrieval~\cite{lee2011shadowdraw, LiuSSLS17, YuLSXHL16} and image editing~\cite{DekelGKLF18,FaceShop18,ArtWF20,DPS20}, simulation control~\cite{hu2019sketch2vf}, block arrangement~\cite{sketch2domino}, and 3D modeling~\cite{Fukusato2020Meshing, igarashi2001suggestive, he2021}. Among sketch-based systems, freehand portrait sketching is difficult for common users due to the  required drawing skills and capabilities, which are inaccessible to novices (e.g., with poor drawing skills). To address this issue, we aim to establish a user-friendly framework to support the process of the freehand drawing of human faces. 

\subsection{Drawing Assistance}
A sketching system's guidance has been thoroughly investigated~\cite{FlaggR06, igarashi1997beautification, LavioleH12}. Especially, displaying visual guidance that can be extracted from reference images (e.g., geometric structures~\cite{DixonPH10, iarussi2013drawing}) on the canvas enables one to support the process of the freeform drawing of objects by tracing over the guidance~\cite{su2014ez, XieHLW14}. However, the user must select reference images, which can be time consuming. Lee et al.~\cite{lee2011shadowdraw} and Choi et al.~\cite{choi2019sketchhelper} dynamically search relevant images from a large-scale database based on intermediate drawing results at drawing time and generate shadow guidance that suggests a sketch completion to users. A similar drawing interface was designed for calligraphy practice~\cite{calli2020}.  With these retrieval-based approaches, the visual guidance may limit in the predefined database. To overcome this issue, image generation approaches can increase the variations from simple strokes, such as Drawfromdrawings~\cite{matsui2017drawfromdrawings} and MaskGAN~\cite{Lee0W020}. Our framework combines both sketch-based retrieval and generation with optimization conversion from sketch-mask mapping.

\subsection{Portraits Rendering}
In the field of non-photorealistic rendering (NPR) of portraits~\cite{Rosin17}, existing approaches typically take one of two approaches. One approach is to extract contour lines from images~\cite{ChenZGLL08,XieZX07,ZhangWX17}. 
While these can be useful for visual abstractions (e.g., preserving and enhancing local shapes), it is difficult to consider semantic constraints and capture specific styles. 
The other approach is to train a network that automatically generates artistic-like drawings from facial images~\cite{LiW16,LiaoYYHK17,SilvaCJM19,ZhuPIE17}. In these problem settings, training a network requires pairs of facial images and portraits. However, it is challenging to construct pixel-based (dense) correspondence because facial components (e.g., eye and nose) in portraits are manually located by artists. Lie et al.~\cite{YiLLR19} combine a global network (for images as a whole) and a local network (for each facial component recognition) and transform high-quality portraits while preserving facial components. 
In this paper, we adopt a similar portrait rendering model to generate portrait drawings, and use them as local guidance.
%

\section{User Interface}
\label{sec:ui}
In this section, we describe how users interact with the proposed two-stage user interface (see Figure~\ref{fig:ui}) to draw portraits with global and local guidance. Please refer to the accompanying video for details. 
%

\subsection{Drawing Tool} 
As with commercial drawing tools, the system enables the user to draw black strokes, in which the stroke width is manually determined using a slider, on canvas with a mouse-drag operation. Then, the system automatically records all the vertices of the strokes and the stroke order for the mask generation step. In contrast, with the eraser tool, the user clicks on a stroke, and the system deletes the selected stroke. Moreover, the undo tool can delete the last stroke from the stroke list. 
Note that our system can also load (or export) the user-drawn strokes by clicking the ``Load'' (or ``Save'') buttons. 

\subsection{Visual Guidance}
Given user-drawn strokes, the system generates two types of visual guidance (i.e., global and local guidance) to use tracing. 
First, in the step of global guidance, the system dynamically searches several relevant images from a database based on the user's intermediate drawing and generates a ``blended'' image (global guidance) rather than a single image. With the global guidance, users can roughly understand locations and shapes of facial parts with correct proportions, as shown in Figure~\ref{fig:ui}(a). 
Second, in the step of local guidance, the system generates several detailed facial portraits (guidance candidates) based on the user's intermediate drawings, and displays ``one'' of them instead of a blended image. The system has a switching function to change the generated images, so the user can search the most reasonable local guidance. By using the local guidance, users can easily design local details such as eyes and nose; see Figure~\ref{fig:ui}(b). 
Note that the system allows the user to freely switch global and local guidance modes by clicking the global/local radio button or the face icon button. 

\hzy{
\subsection{Rewind Tool}
In order to help users to draw the desired portraits, we provide the rewind tool in the proposed user interface. If users thought the local guidance does not meet their vision, the drawing process can return to the global stage by selecting the corresponding radio button, as shown in Figure 2. Our drawing interface can automatically save the sketches while switching among global and local stages, so that users can revise their drawn contour sketches by reloading the recorded data.}

\section{Two-Stage Drawing Guidance}
\label{sec:method}
Inspired by conventional portrait drawing processes, this work proposes dualFace, a two-stage framework for portrait drawing with both a global stage and a local stage for drawing guidance. For the global stage of user guidance, we provide interactive drawing guidance for each facial part. To help users achieve balanced facial contour drawing, we adopt the data-driven facial feature query by matching the Gabor Local Line-based Feature (GALIF)~\cite{EitzRBHA12}. For the local stage of user guidance, we adopt a GAN-based neural network to generate corresponding fine-grained sketches from a user’s rough contour sketch in the global stage. Since we provide the photo-realistic facial details in the local guidance, dualFace can help users concentrate on detailed drawing for facial features and improve their drawing skills. We believe the two-stage framework of dualFace may narrow the gap between novices and artists on portrait sketching due to the separation of the global contour information from local facial details.

\begin{figure*}[t]
\includegraphics[width=1.0\textwidth]{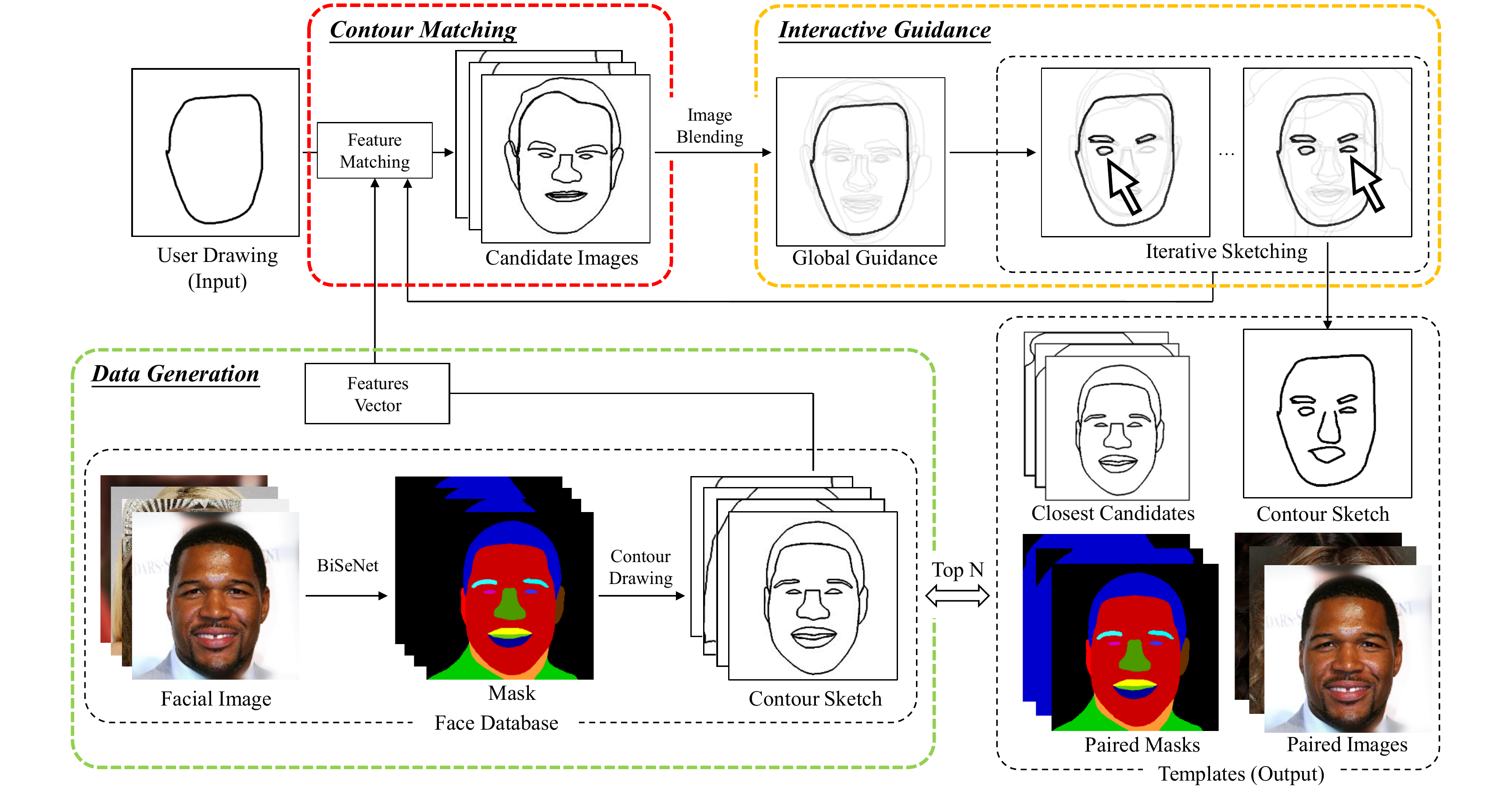}
\caption{The stage of global guidance consists of three steps: data generation, contour matching, and interactive guidance. 
\hzyv{The contour sketches in our database are extracted from masks as source images which are more meaningful for feature matching to achieve better drawing guidance than previous work~\cite{lee2011shadowdraw}.}
}
\label{fig:global}
\end{figure*}

\subsection{Global Guidance}

It is difficult to draw recognizable portraits with correct locations and portions of facial features, especially, for novices. To solve this issue, dualFace first aims to help users to draw balanced facial contours. Figure \ref{fig:global} shows the workflow of global guidance, including data generation, contour matching, and interactive guidance. For the data generation step, face images are converted to contour images from a face database. For the contour matching step, the local facial features are calculated and stored as feature vectors indexed into the database. For the interactive guidance step, the most similar candidates are retrieved as the shadow guidance in real time. \hzy{In contrast to the previous work of Shadowdraw~\cite{lee2011shadowdraw} with edge maps, we adopted the labelled contour sketches for feature matching with the semantic sketch information. Therefore, each stroke of users’ drawing input can be matched with the meaningful facial features for the next stage of local guidance.}

\subsubsection{Data Generation}  
It is challenging to collect an enormous number of artist-designed portraits for face retrieval. Instead of the artistic portraits, we generate semantic label masks~\cite{YuWPGYS18} by utilizing a bilateral segmentation network (BiSeNet) pretrained on the CelebAMask-HQ dataset~\cite{Lee0W020}. Each pixel in the masks has a facial label ID from facial images (e.g., eyes, nose, and mouth). We adopted the contour function of OpenCV library for the line drawing functions. The contours of facial components are extracted from the semantic label masks with the balanced facial features. 
Note that the contour images are stored with the corresponding original face images, which are used for sketch retrieval on the global stage and for system input on the local stage. 

\subsubsection{Contour Matching}
To explore the closest contour sketches from the database as the guidance according to a user's incomplete freehand sketch  in real time, we use GALIF features for sketch retrieval and local shape matching~\cite{EitzRBHA12}. 
For the online query method, the user sketch is encoded as a histogram. We calculate the similarity with the stored contour images in our database to obtain the closest contour images.

\subsubsection{Interactive Guidance}
Similar to the shadow drawing interface~\cite{lee2011shadowdraw}, the top $N$ relevant retrieval results in the face database are merged as a shadow image by image blending ($N = 3$ in our implementation). Benefiting from the interactive global guidance for portrait drawing, users could realize the locations and shapes of each facial part. The global guidance is updated in real time for each drawing stroke. Under the help of global guidance for portrait sketching, the user can complete the contour sketch to express the rough shape and the location of facial parts meeting their drawing intentions.

\begin{figure*}
\includegraphics[width=1.0\textwidth]{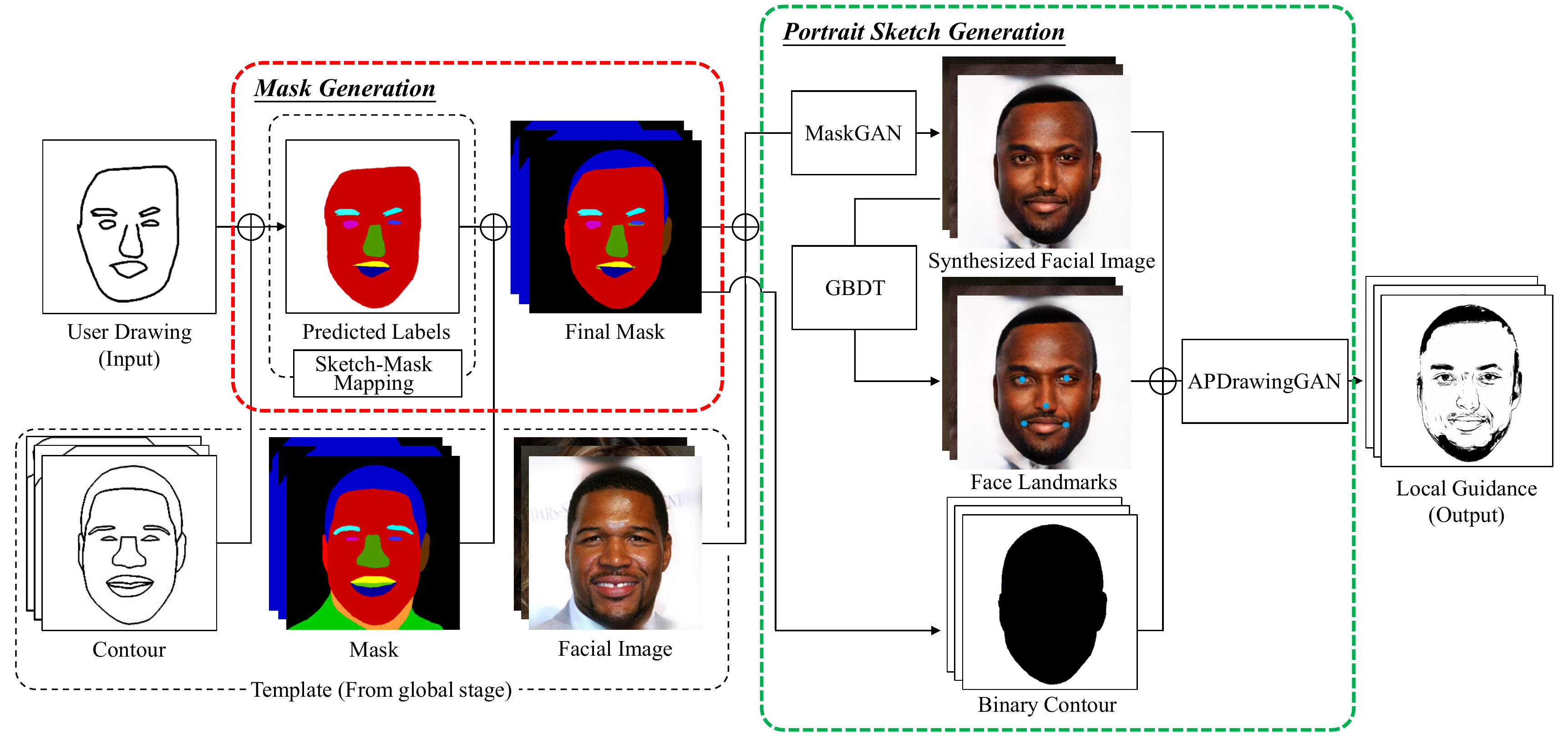}
\caption{
The stage of local guidance consists of two steps: mask generation and portrait sketch generation. 
}
\label{fig:local}
\end{figure*}

\subsection{Local Guidance} 
In order to guide users to draw details of facial components (e.g., black irises and eyelashes), dualFace provides the local guidance using relevant templates extracted from our database in the global stage. Local guidance for portrait sketching includes mask generation and portrait sketch generation (Figure~\ref{fig:local}). 
For the mask generation step, user strokes in the global stage are recorded and converted to face masks based on the top $N$ relevant the templates ($N=3$ in our implementation). For portrait sketch generation step, all templates can generate fine-grained portrait sketches, and the user can select the most desirable one as the reference for further drawing. Note that the input contour sketch is not required to contain all facial parts, and the missing parts can be completed automatically with our stroke-mask mapping optimization.

GAN-based neural networks are used in local guidance for mask and portrait sketch generation. In our implementation, we adopted MaskGAN~\cite{Lee0W020} to generate portrait images matching the facial contour sketch and APdrawingGAN~\cite{YiLLR19} to transfer the portrait images into artistic portrait sketching. Note that two generative models are trained independently. To connect two models, the facial landmarks are calculated with Gradient Boosting Decision Tree (GBDT)~\cite{KazemiS14}, and the binary background mask is converted from the merged mask.

\subsubsection{Mask Generation} 
\hzy{For portrait image generation, the conventional approaches adopted the facial mask with manually defined label information as shown in Figure 4 (red dash line of mask generation, and different colors denote facial labels). However, it is a boring and a time consuming task of manual labeling for portrait drawing in our work. To alleviate the manual labors and adapt to freehand sketching, we proposed automatic sketch-mask mapping with an optimization algorithm to generate facial mask according to the contour sketch from user drawing.}

We first calculate the shape similarity $F$ between user-drawn strokes $S$ and regions of face template mask $M$. Any single stroke $\bm s \in S$ can be regarded as in-sequence vertices, where $\bm s=\{\bm p_{i}\,|\,i = 1, \cdots, N\}$. Then, we obtain the correspondence between two regions using the following equation.
\begin{equation}
\begin{split}
F(S, M)&=\min_{\bm s} \sum_{\bm s \in S}{Dis(\bm s, m_k)} \\
&=\min_{\bm p} \sum_{\bm s \in S}(\frac{1}{N}\sum_{\bm p \in \bm s}{L_2(\bm p, m_k)})\\ 
& \!\!\!\!\!\!\!\!\!\!s.t.\: label(\bm s)=k\ and\ m_k \in M
\end{split}
\end{equation}
\noindent
where $Dis(\bm s, m_k)$ denotes the distance between a single stroke $\bm s$ with $m_k$ (region of $M$ with the label ID is $k$). $Dis(\bm s, m_k)$ consists of $dis(\bm p, M)$, which denotes the average of $L_2$ distance from all vertices $\bm p \in s$ to $m_k$. $label(\bm s)$ is the discriminant function to calculate the label ID of $\bm s$ decided by the majority vote algorithm of vertex $\bm p \in \bm s$, as calculated by the following equations:

\begin{equation}
\label{vvote}
\left\{\begin{matrix}
 label(s)=\argmax_{\bm p} C_{\bm p \in s}{(V(\bm p, M))}&  & \\ 
 V(\bm p, M)=k^*=\argmin_{\bm p}\ dis(\bm p, m_k)
\end{matrix}\right.
\end{equation}

\noindent where $C_{\bm p \in \bm s}(\cdot)$ is the aggregate function for stroke $s$ to count the number of its vertices with the same label ID. 
Discriminant function $V(\bm p, M)$ can determine the label ID $k^*$ for a single vertex $\bm p$ in $M$ by searching the minimum distance of $\bm p$ in each region of $M$.

The sketch-mask mapping algorithm is described in Algorithm \ref{algo:smmap}. User’s strokes are classified to the labels in the matching mask respectively, and strokes with the same labels are merged as a new stroke. Then, a contour (concave hull) of each new stroke is calculated as a new mask to replace the old one in the matching mask.

\begin{algorithm}[t]
	\caption{Sketch-Mask Mapping}
	\label{algo:smmap}
	\KwIn{Strokes $S$, Matched mask $M$}
	\KwOut{User-defined mask $M^*$}
	$M^* =zeros( M$.shape)\; 
	Number of mask $M$ $m \leftarrow len(M)$\; 

	\For{k=1:m}
	{
		Merged stroke with same label $ms$\;
		Mask region in same label $mask = M[k]$\;
		\If{mask is None}
		{
			continue;
		}
		$ms$= MergeStrokes($s \in S$ s.t. $ label(s)==k$) \;
		\If{ms is None}
		{
			$M^*[k]=mask$;
		}
		$M^*[k]$=ConcaveHull($ms$)\;
	}
	return $M^*$

\end{algorithm}
In terms of the correspondence between the user sketch and face template, we transfer semantic labels of facial components in the facial template to each region of the user-drawn stroke (e.g., hair, mouth, eyes). Then, we replace the corresponding template regions with ones of user-drawn regions if existed and merge the user's stroke feature into the mask. Note that the contour sketch can be auto-completed even if the user input sketch is partial. Finally, we replace user-drawn regions (partial sketch) and the corresponding template regions, and generate a complete label mask.

\subsubsection{Portrait Sketch Generation}
Generating facial images with details from rough sketches is an under-determined problem. 
An end-to-end GAN-based model requires extensive artistic drawing with similar styles for training, which is expensive and time consuming. To solve this issue, we divide this problem into sketch-to-portrait image generation and artistic rendering for simplification as shown in Figure~\ref{fig:local}. We first generate a realistic facial image using the MaskGAN network based on the complete label mask, corresponding face image, and the face template from the global stage. Then, we convert the face image to a portrait sketch using the APDrawingGAN network for artistic rendering . We obtain the locations of facial components based on GBDT and binary contour of the background from the final mask to connect the two generative networks of mask and portrait sketch generation. \hzy{Note that the global features of the generated local references have been restricted by users’ contour sketch.}

\subsection{Implementation}
In our implementation, dualFace was programmed in Python as a real-time drawing application on the Windows 10 platform. 
A workstation with Intel Core i9 10900KF, 3.7GHz 5.10GHz, NVIDIA RTX2080ti GPU $\times$ 2, and 64GB RAM was used as the testing computing environment. 
In addition, 518 images with a size of 512 $\times$ 512 were picked up from the CelebAMask-HQ dataset and converted to contour sketches. GALIF features were extracted for sketch retrieval on the stage of global guidance. For the implementation of local guidance, we used MaskGAN for mask generation consisting of the Dense Mapping Network (DMN) for image generation and U-Net like MaskVAE for mask editing, which is pre-trained on CelebAMask-HQ with more than 200 thousands images. We used APDrawingGAN for portrait sketch generation with a hierarchical GAN structure using U-Net with skip connections for each facial feature (i.e., left eye, right eye, nose, and mouth). In this work, we utilized the pretrained models with 300 epoch training on the APDrawing dataset (140 face images and corresponding portrait drawings by an artist). 

Our prototype system requires, on averagely, 0.36s for image retrieval in global guidance after mouse release every time and 2.78s for each portrait image generation in local guidance. Note that the image generation was conducted only once, meaning dualFace can provide effective feedback for portrait drawing. \hzyv{Because dualFace generates facial images for local guidance, there is no reference images or labels available as ground truth for quantitative evaluation. Therefore, we conducted user study to verify our approach in a qualitative way.}
\begin{table*}[htbp]
\centering
\caption{Questionnaire results in our user study.}
\vspace{1mm}
\begin{tabular}{c|l|c|c|c}
\hline
\# & \multicolumn{1}{c|}{Question} & Score & Mean & SD \\\hline\hline
\multirow{3}{*}{\raisebox{-1mm}{\rotatebox{90}{Global}}} & \raisebox{-0.3mm}{Clear and easy to understand?} &\multirow{9}{*}{
\begin{minipage}{48mm}
    \centering
    \scalebox{1.005}{\raisebox{-0.3mm}{\includegraphics[width=1.0\textwidth]{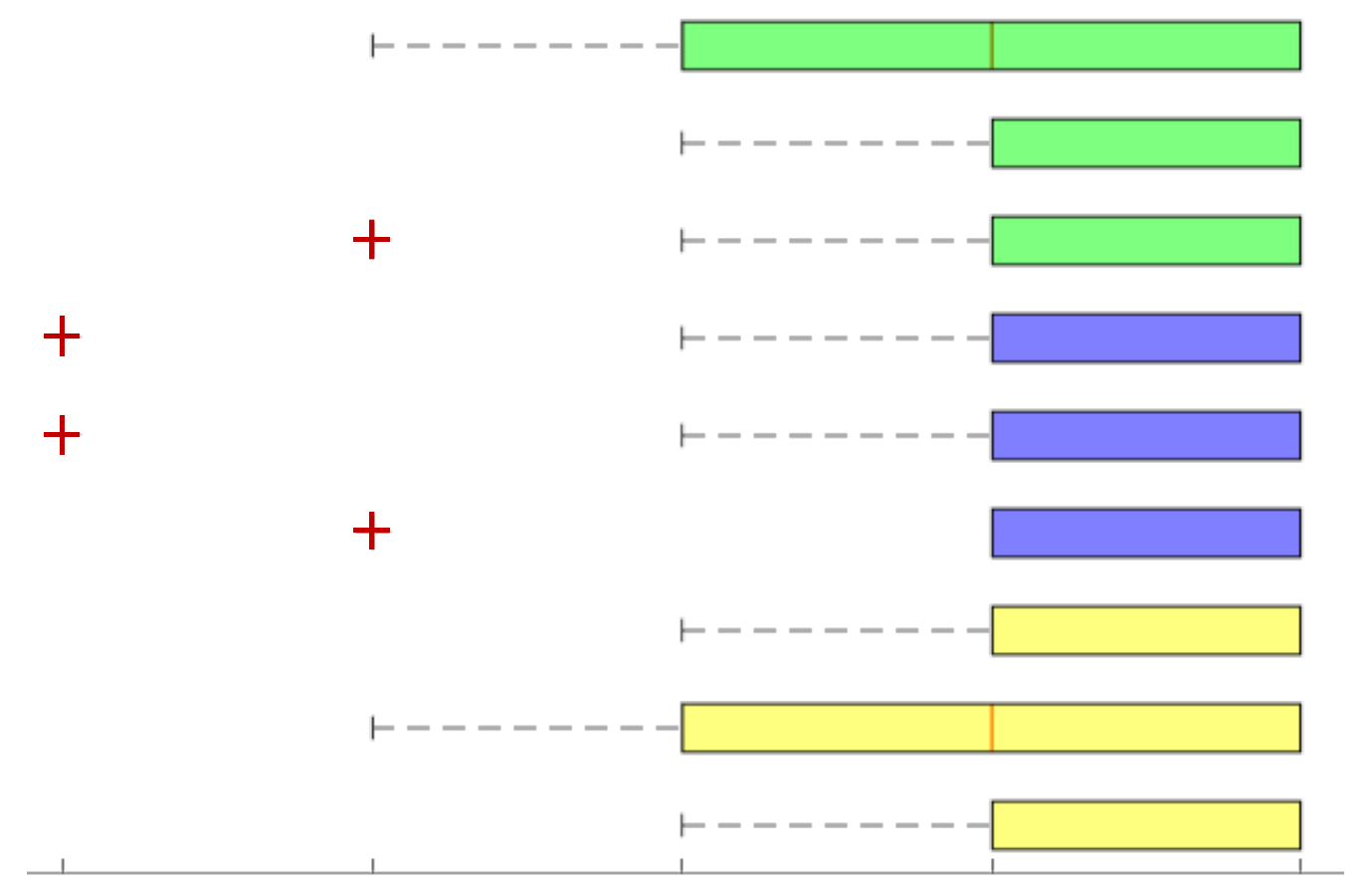}}}
\end{minipage} 
}
& \raisebox{-0.3mm}{4.5} & \raisebox{-0.3mm}{0.6} \\
 & \raisebox{-0.3mm}{Feedback is meaningful and helpful?} && \raisebox{-0.3mm}{4.1} & \raisebox{-0.3mm}{1.0} \\
 & \raisebox{-0.3mm}{Easy to follow and use?} && \raisebox{-0.3mm}{4.1} & \raisebox{-0.3mm}{0.8} \\ \hline
\multirow{3}{*}{\raisebox{-1mm}{\rotatebox{90}{Local}}} & \raisebox{-0.3mm}{Clear and easy to understand?} && \raisebox{-0.3mm}{4.6} & \raisebox{-0.3mm}{0.8} \\
 & \raisebox{-0.3mm}{Feedback is meaningful and helpful?} && \raisebox{-0.3mm}{4.0} & \raisebox{-0.3mm}{1.1} \\
 & \raisebox{-0.3mm}{Easy to follow and use?} && \raisebox{-0.3mm}{4.0} & \raisebox{-0.3mm}{1.0} \\ \hline
\multirow{3}{*}{\raisebox{-1mm}{\rotatebox{90}{Overall}}} & \raisebox{-0.3mm}{Helped me learn how to draw faces?} && \raisebox{-0.3mm}{4.2} & \raisebox{-0.3mm}{0.9} \\
 & \raisebox{-0.3mm}{Useful for helping learn how to draw faces?} && \raisebox{-0.3mm}{4.1} & \raisebox{-0.3mm}{0.6} \\
 & \raisebox{-0.3mm}{Useful for helping improve face drawing skill?} && \raisebox{-0.3mm}{3.9} & \raisebox{-0.3mm}{1.2} \\ \hline
\end{tabular}
\label{fig:question}
\end{table*}

\section{User Study}
\label{sec:study}

To evaluate the usefulness of the proposed user interface~(UI) dualFace, we compared dualFace with two conventional drawing interfaces: suggestive drawing UI (Figure~\ref{fig:question}(a)) and shadow drawing UI (Figure~\ref{fig:question}(b)). The implemented suggestive drawing UI provides the three most related contours in sub-windows below the main canvas from the face database. The shadow drawing UI provides the blended shadow image from dualFace's global stage, similar to Shadowdraw~\cite{lee2011shadowdraw}. 

\begin{figure}[htbp]
\centering
\includegraphics[width=1.0\linewidth]{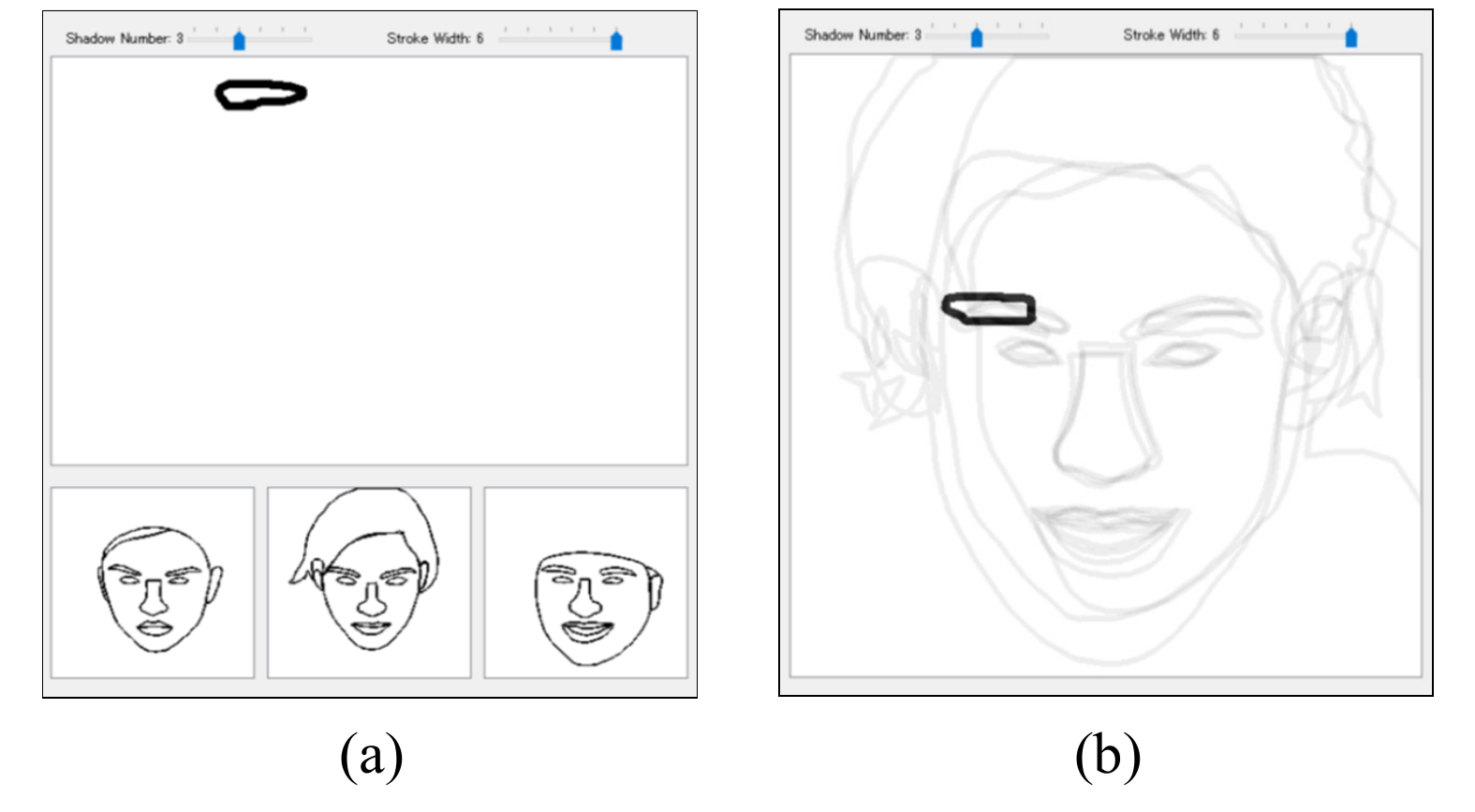}
\caption{Drawing interfaces used in our user study: (a) suggestive drawing UI and (b)~shadow drawing UI.}
\label{fig:compareUI}
\end{figure}

\subsection{Evaluation Procedure}
We invited 14 participants in our comparison study (graduate students, nine males and five females). All participants were asked to draw portraits with a pen tablet (WACOM with 22.4 cm $\times$ 14.0 cm drawing area) and a LCD monitor (126.2 cm$\times$83.7 cm). All participants were asked to draw \hzyv{freely and aimlessly and try to draw more details as possible as they can} with all three drawing interfaces: suggestive UI, shadow UI, and ours \hzy{in random order}. They first drew freely on the tablet until they felt comfortable using the devices before the user study. We instructed all participants how to use dualFace with a usage manual. Considering the usage of facial masks, we asked the participants to draw each facial mask in a well-closed curve. All participants were required to draw carefully and choose the most anticipated references for local guidance \hzy{from multiple generated candidates} after they completed the global stage. Finally, we administered the questionnaire to all participants after they finished the user study.

The questions in our questionnaire were designed to confirm the effectiveness of global and local guidance, and the overall evaluation using dualFace, as shown in Table~\ref{fig:question}. All questions adopted five-point Likert scale (1 for strongly disagree, 5 for strongly agree). 

\subsection{Drawing Evaluation}
After all participants completed the comparison study, the other 25 participants joined the online questionnaire for drawing quality evaluation. All participants were asked to score up 12 portrait sketches (four for each drawing UI). We confirmed two questions about the qualities of the spatial relationship and facial details for all portrait sketches. We adopted five-point Likert scales for all questions (1 for very poor, 5 for very good). A good spatial relationship of portrait sketch means the well-balanced facial parts, and good facial details mean that each facial part has finely detailed drawing, such eyes and mouth. We explained the meanings of the two qualities to all participants before the online questionnaire.

\section{Results}
\label{sec:result}
We discuss the implementation results of dualFace, evaluation results, user feedback and our observations from our user study.


\newcommand\wsize{0.12}

\begin{figure*}[t]
\begin{minipage}{0.2\linewidth}\centering
\vspace{\vcsize}
\begin{spacing}{5.0}
User sketch\\
Revised contour\\
Local guidance\\
Final result
\end{spacing}
\vspace{-\vcsize}
\end{minipage}
\renewcommand\userid{1}
\begin{minipage}{\wsize\linewidth}\centering
\includegraphics[width=\textwidth]{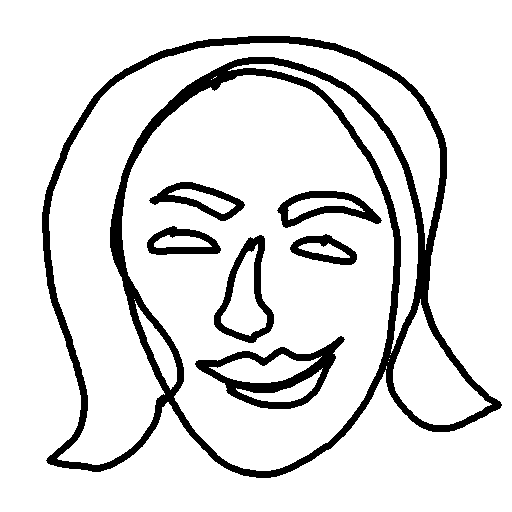}\\
\includegraphics[width=\textwidth]{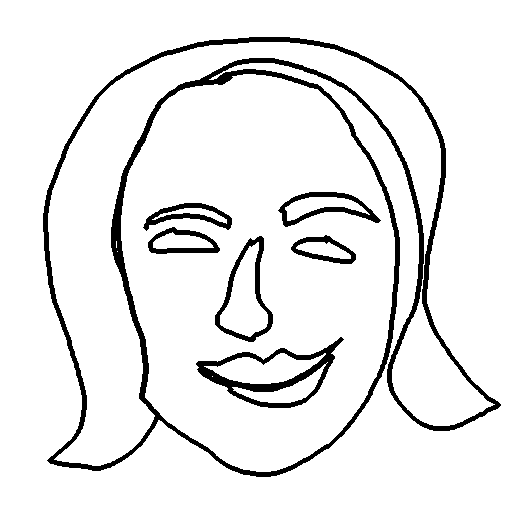}\\
\includegraphics[width=\textwidth]{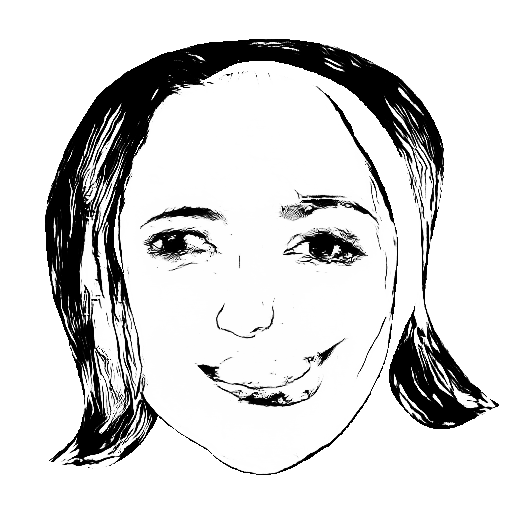}\\
\includegraphics[width=\textwidth]{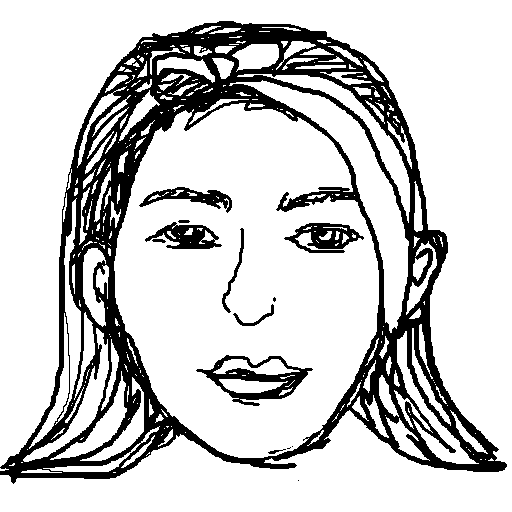}\\
\end{minipage}
\renewcommand\userid{2}
\begin{minipage}{\wsize\linewidth}\centering
\includegraphics[width=\textwidth]{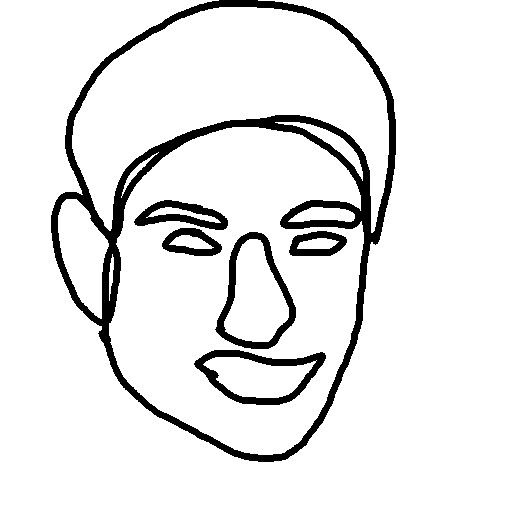}\\
\includegraphics[width=\textwidth]{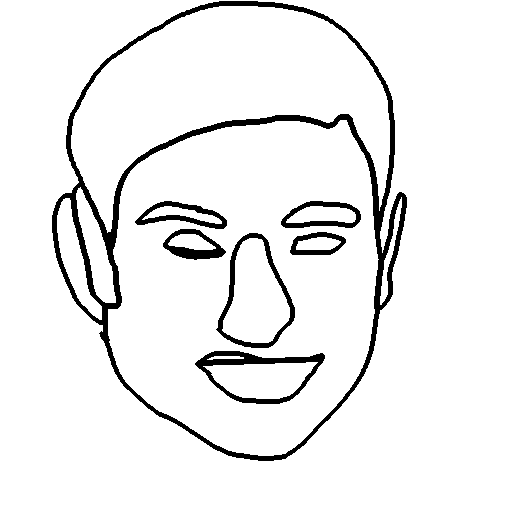}\\
\includegraphics[width=\textwidth]{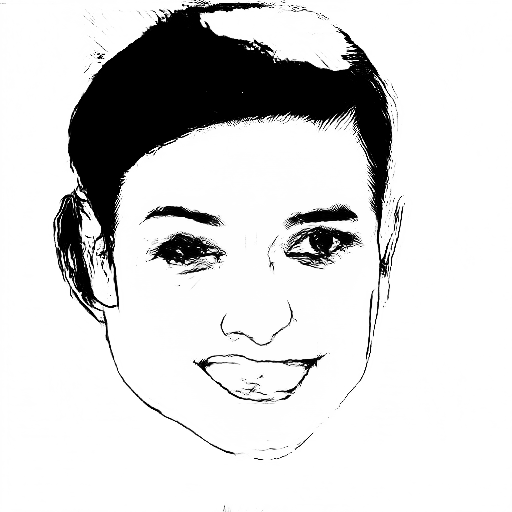}\\
\includegraphics[width=\textwidth]{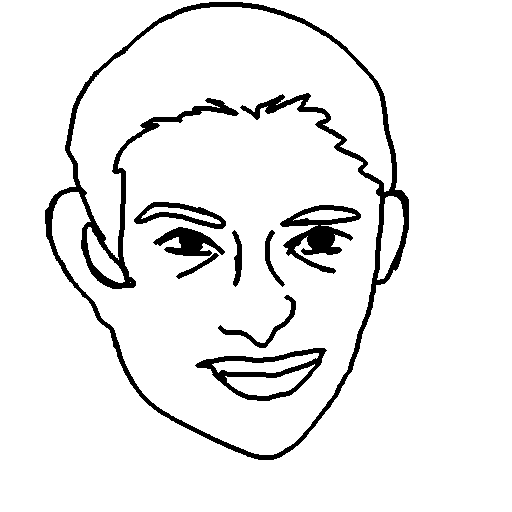}\\
\end{minipage}
\renewcommand\userid{4}\begin{minipage}{\wsize\linewidth}\centering
\includegraphics[width=\textwidth]{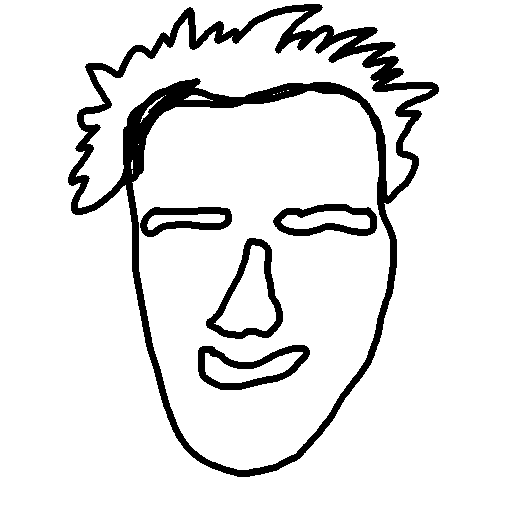}\\
\includegraphics[width=\textwidth]{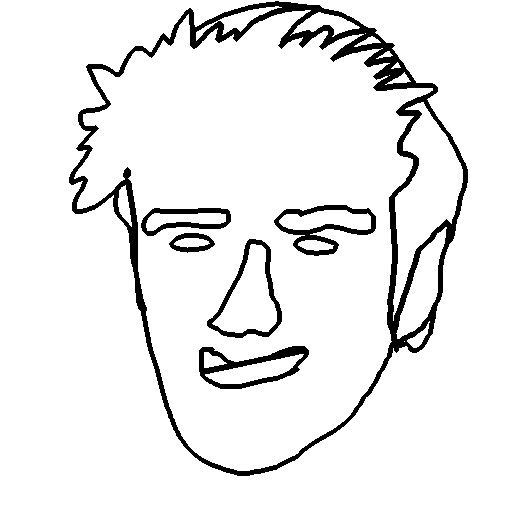}\\
\includegraphics[width=\textwidth]{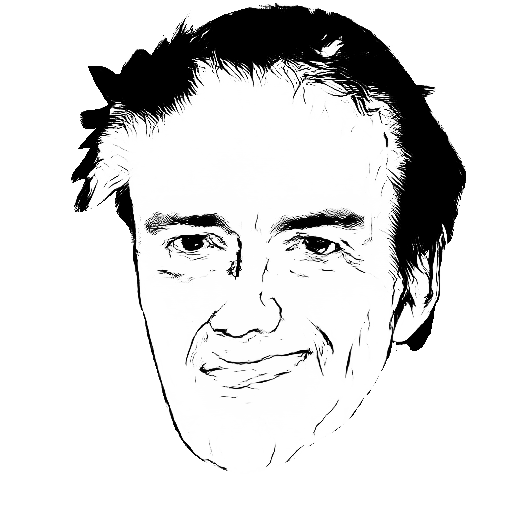}\\
\includegraphics[width=\textwidth]{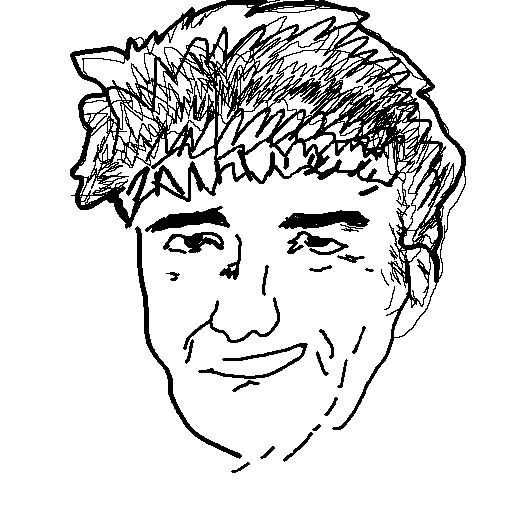}\\
\end{minipage}
\renewcommand\userid{5}\begin{minipage}{\wsize\linewidth}\centering
\includegraphics[width=\textwidth]{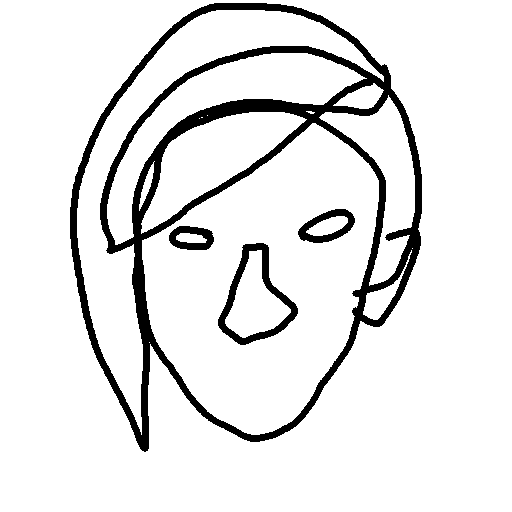}\\
\includegraphics[width=\textwidth]{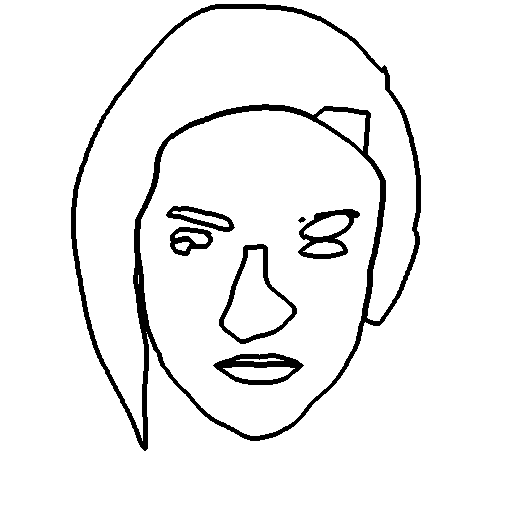}\\
\includegraphics[width=\textwidth]{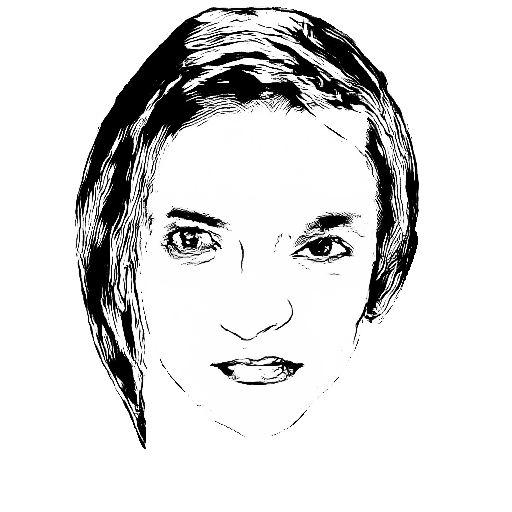}\\
\includegraphics[width=\textwidth]{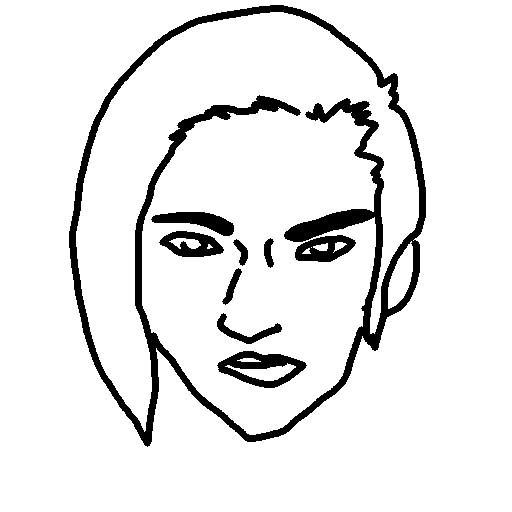}\\
\end{minipage}
\renewcommand\userid{6}\begin{minipage}{\wsize\linewidth}\centering
\includegraphics[width=\textwidth]{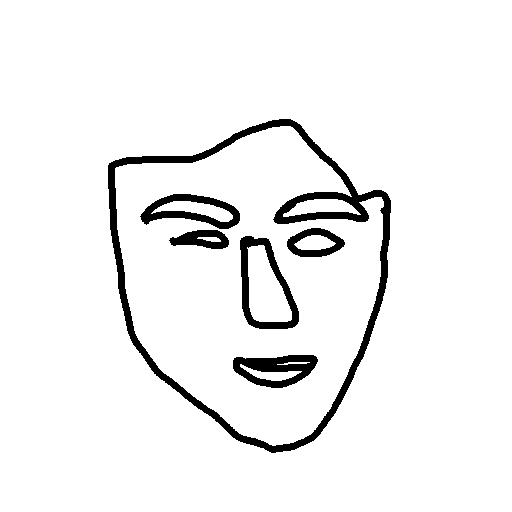}\\
\includegraphics[width=\textwidth]{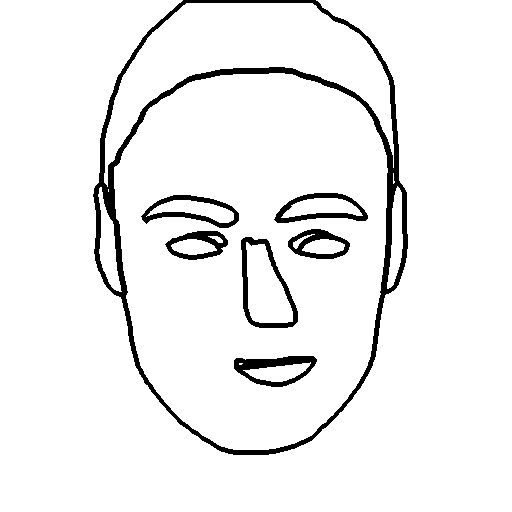}\\
\includegraphics[width=\textwidth]{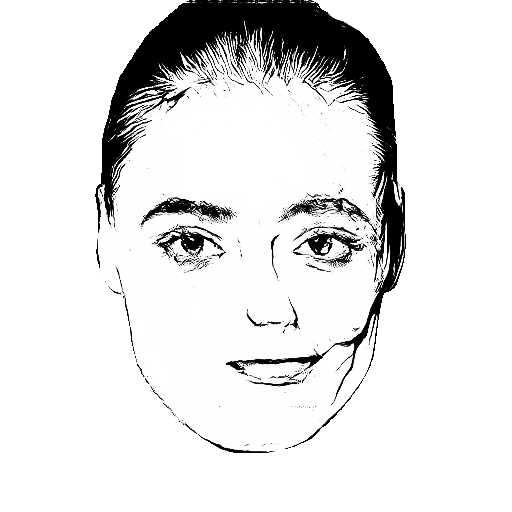}\\
\includegraphics[width=\textwidth]{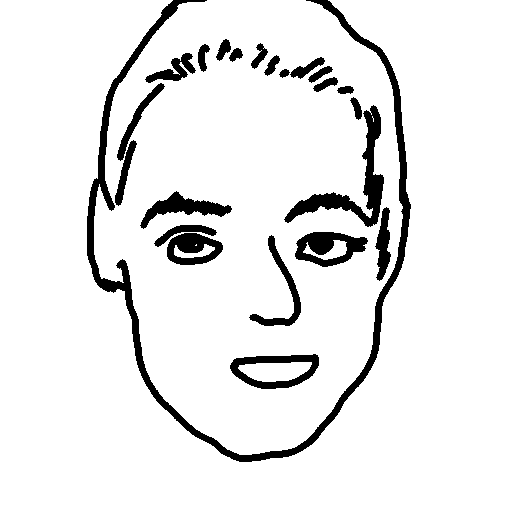}\\
\end{minipage}
\renewcommand\userid{3}
\begin{minipage}{\wsize\linewidth}\centering
\includegraphics[width=\textwidth]{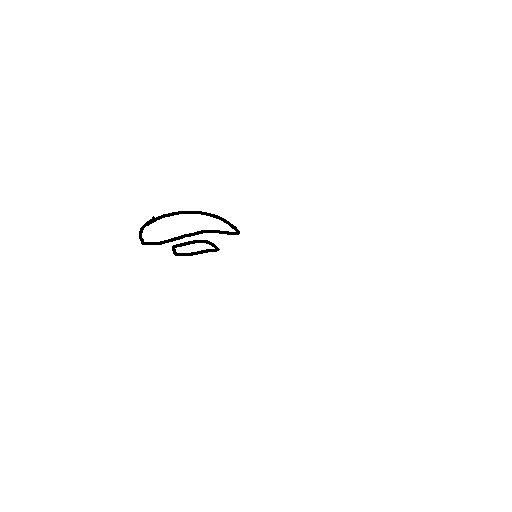}\\
\includegraphics[width=\textwidth]{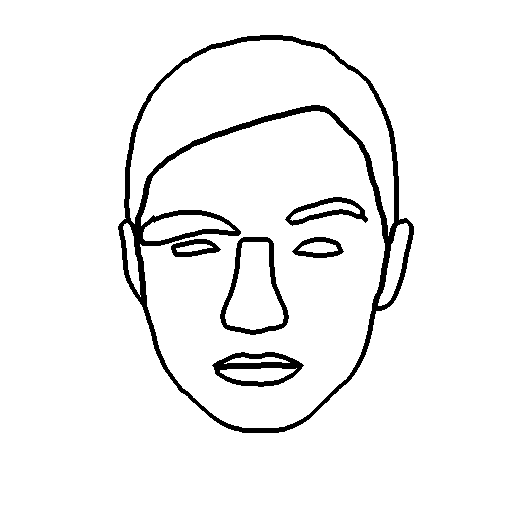}\\
\includegraphics[width=\textwidth]{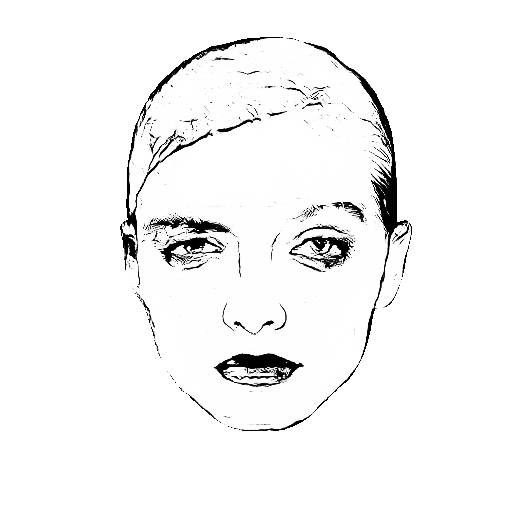}\\
\includegraphics[width=\textwidth]{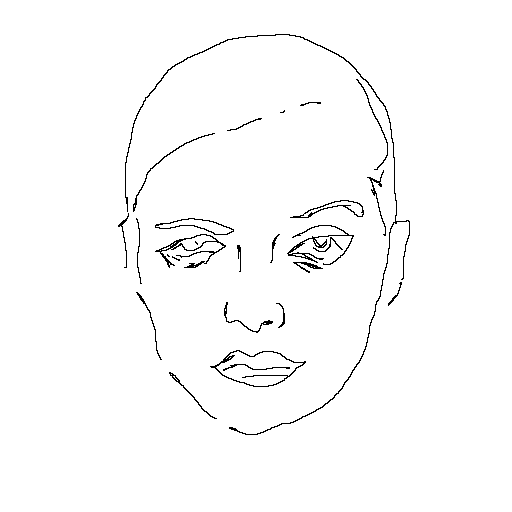}\\
\end{minipage}
\vspace{2mm}
\caption{
Some examples of our implementation results. \textit{User sketch} denotes results drawn under the global guidance. \textit{Revised contour} denotes the matched facial masks in local guidance. \textit{Local guidance} denotes the generated portrait sketch image for reference to the user. \hzy{Note that the portrait images in local guidance were selected by users as closest alternatives to user drawing expectations}. \textit{Final result} denotes the final outcome from user drawing. 
}
\label{fig:y1}
\end{figure*}

\begin{figure}[htb]
\centering
\includegraphics[width=0.85\linewidth]{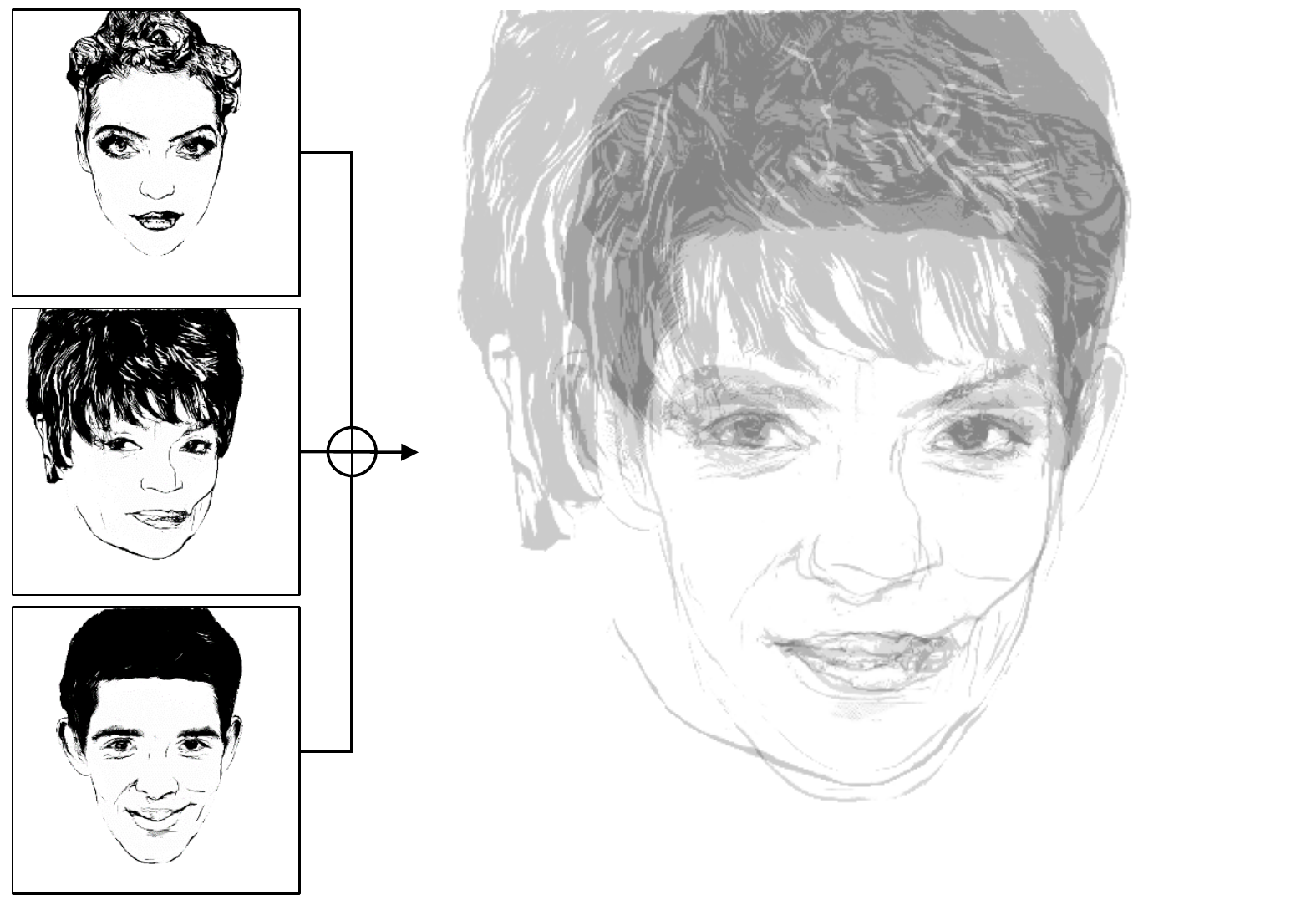}
\caption{As a limitation of ShadowDraw, the blended image (right) has difficulty preserving details of facial images (left).}
\label{fig:imp4}
\end{figure}

\subsection{Visual Guidance} 

Figure~\ref{fig:y1} shows some examples of our implementation results sketched with dualFace. Users can achieve the desirable local guidance according to their freehand contour sketches from the global guidance. If a user's sketch is incomplete, it can be completed automatically and revised with our sketch-mask matching optimization. The last column of Figure~\ref{fig:y1} shows an example of a partial sketch. Although the user only has drawn the left eye and eyebrow contour sketch on the global stage, the proposed system can still work well.

Compared with previous work of the drawing interface ShadowDraw~\cite{lee2011shadowdraw}, dualFace has no limitation on facial details in drawing guidance. If we blend the relevant templates (face images with details), it is difficult to distinguish the facial references with the loss of facial details, as shown in Figure~\ref{fig:imp4}. Therefore, ShadowDraw can only support the drawing guidance of simple subjects without photo-realistic details.

\begin{figure}[t]
\centering
\includegraphics[width=0.97\linewidth]{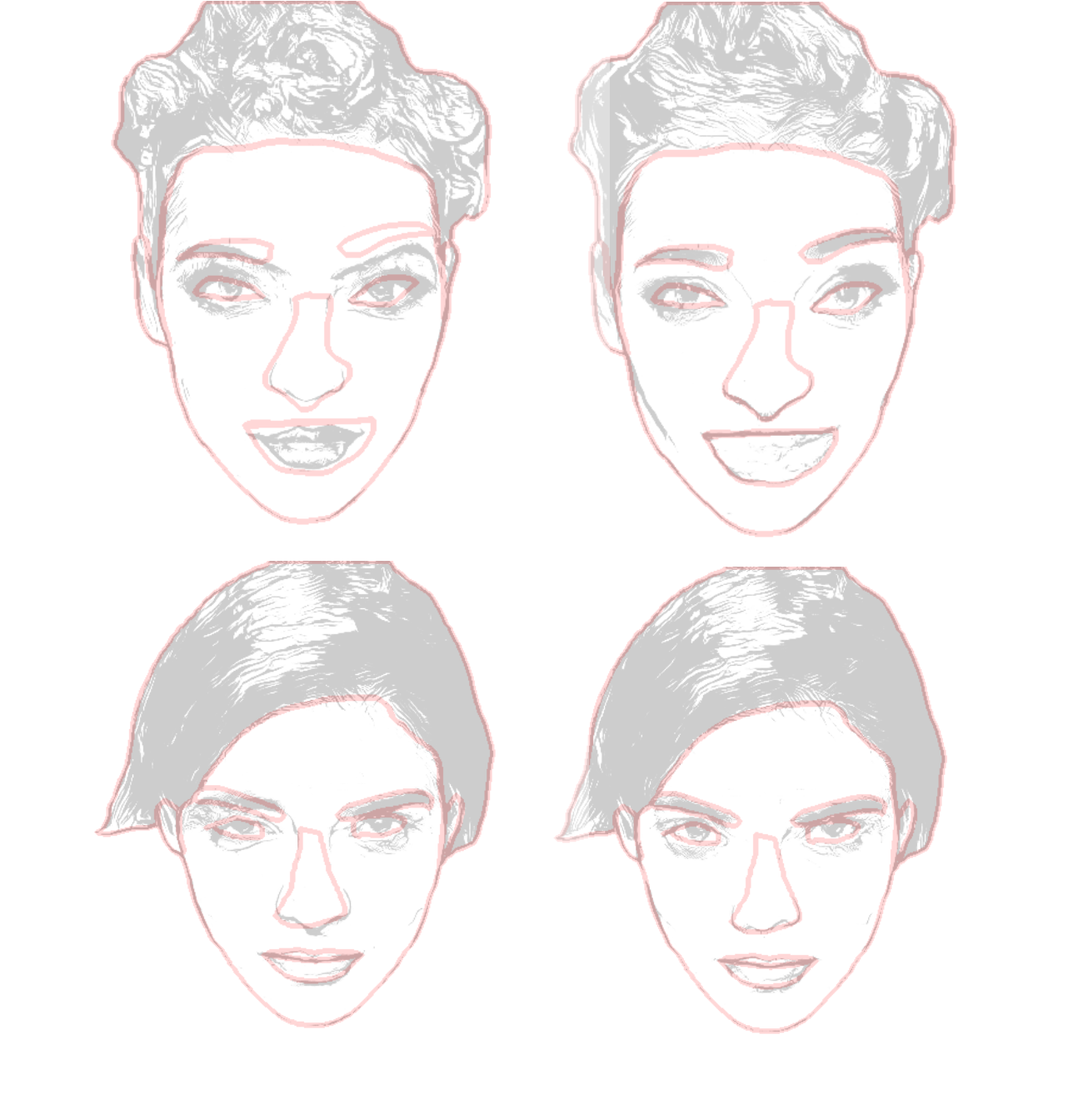}\\
Without mask generation \quad  With mask generation (ours)
\caption{
Comparison results with and without the mask generation process. Mismatches are obvious between the user's contour sketch (red lines) and the generated local guidance without mask generation (left).
}
\label{fig:imp2}
\end{figure}

In local guidance, mask generation plays an important role in meeting the user's intention in freehand drawing. To verify this issue, we compared the system results with and without mask generation, as shown in Figure~\ref{fig:imp2}. In the case without the mask generation process, the feature lines in the user's contour sketch did not conform to the generated portrait drawing, as shown in Figure~\ref{fig:imp2} (left). Meanwhile, more plausible results were achieved by the proposed framework with mask generation. 

\subsection{User Evaluation} 

\begin{figure}[htbp]
\centering
\includegraphics[width=0.98\linewidth]{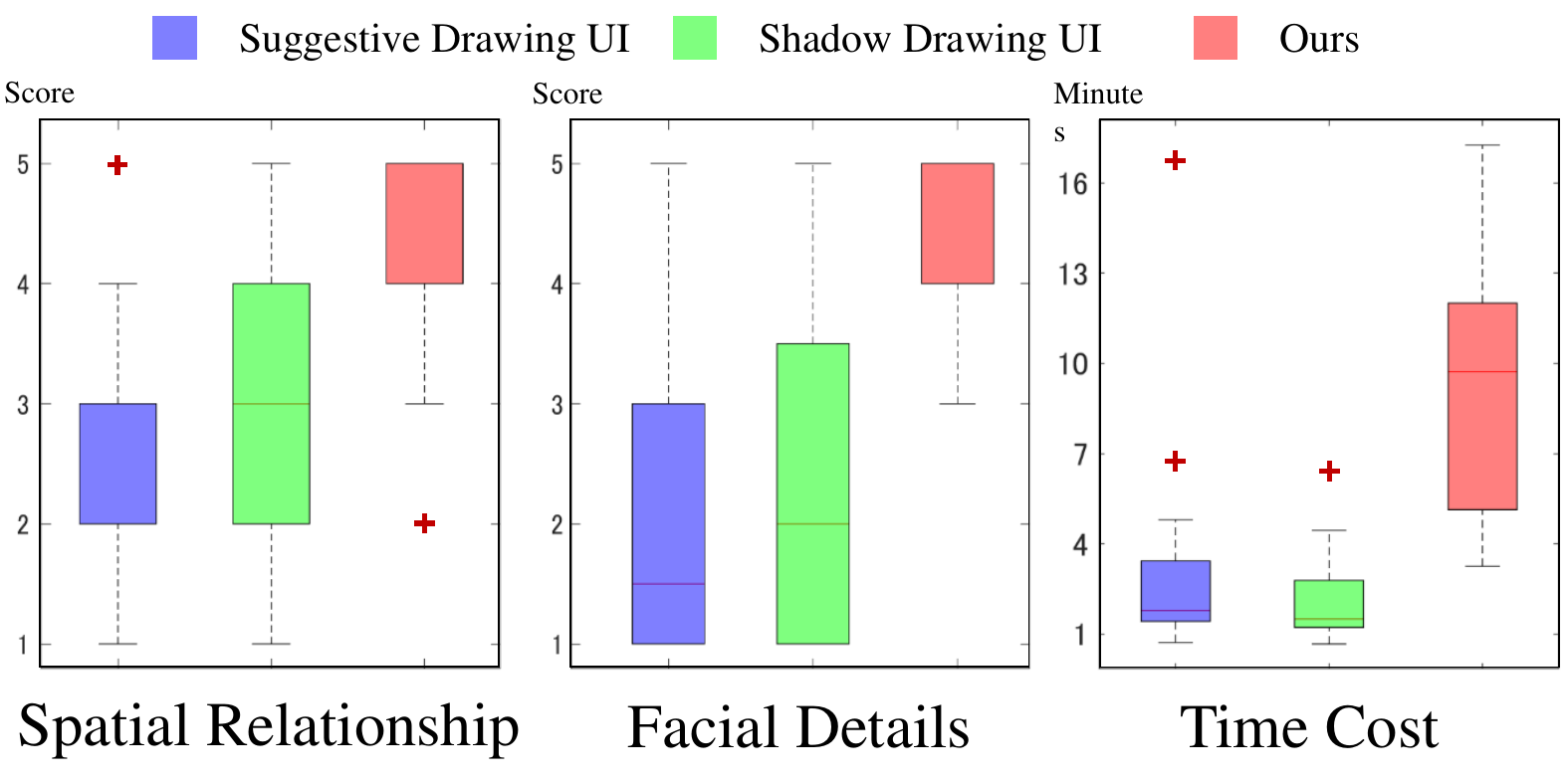}
\caption{Evaluation results of spatial relationship and facial details in portrait drawings (left and middle). Time cost to complete portrait sketching for each drawing interface (right). }
\label{fig:time}
\end{figure}

The results of questionnaire are illustrated in Table~\ref{fig:question}. Participants were asked to score dualFace by answering nine questions in total (three for global guidance, three for local guidance, and three for overall evaluation). The mean scores of all questions are above 3.9, verifying that the proposed drawing interface dualFace is easy to understand and follows at a high level. For overall user experiences, all participants thought our system can help them to draw portrait well and improve their drawing skills. Because dualFace provides guidance on a whole portrait sketch to the participants, users may want to practice the basic drawing skills such as arrangements of straight lines or curves. We plan to improve the current drawing interface to help users practice basic drawing skills in near future.

\begin{figure*}[htbp]
\centering
\includegraphics[width=0.98\linewidth]{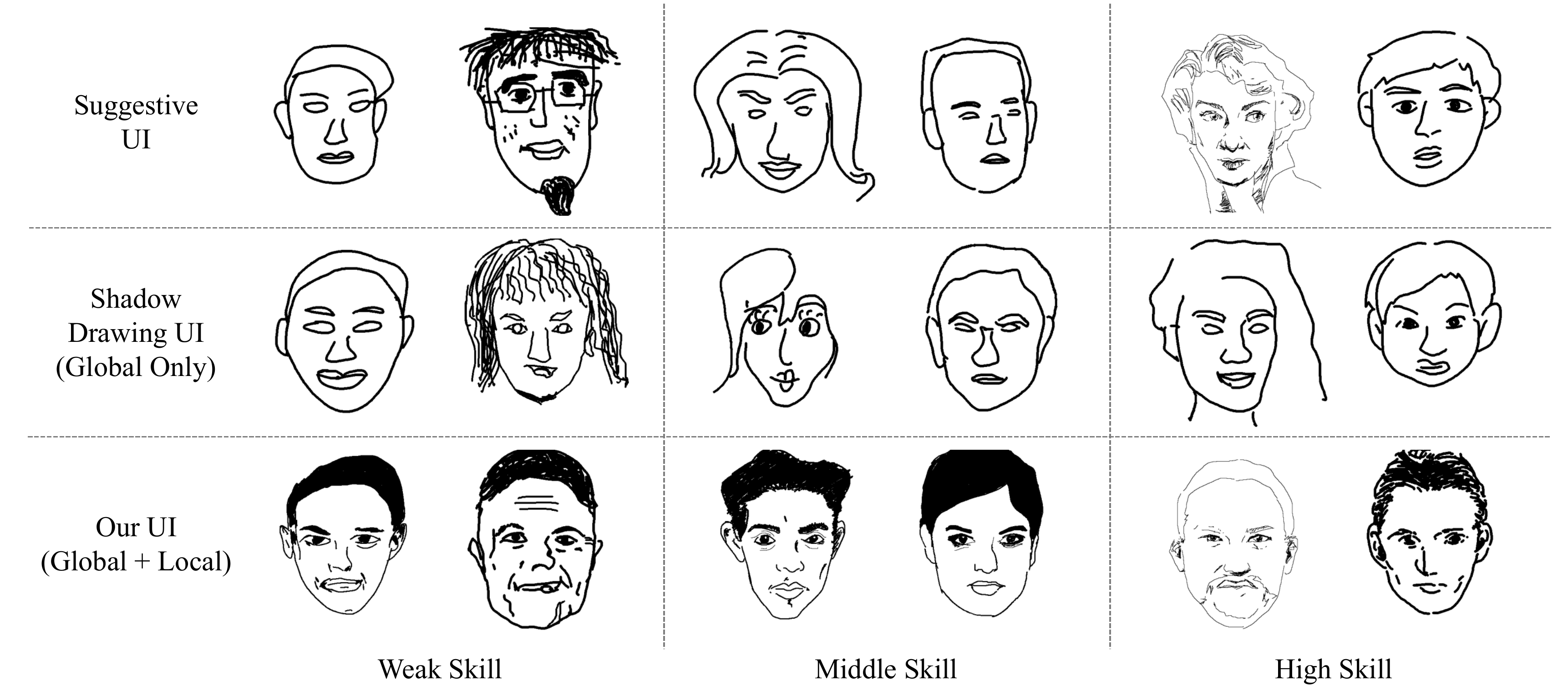}
\caption{Drawing results from six participants. Each column corresponds to the same participant's drawing. }
\label{fig:result}
\end{figure*}

Figure~\ref{fig:time} shows the results of an evaluation study of portrait sketches from our online questionnaire. The proposed drawing interface achieved comparatively high scores in drawing evaluations of both spatial relationship and facial details, and the average scores are 4.5 and 4.32, respectively. Therefore, dualFace can guide users to achieve better portrait drawings with correct facial spatial relationships and detailed facial features, whereas the other drawing interfaces may fail to provide them.

Figure~\ref{fig:result} shows the portrait sketches from our comparison study among suggestive drawing UI, shadow drawing UI, and our dualFace UI in our comparison study. We found that dualFace can not only help users with weak or middle drawing skills to achieve much better portrait sketches, but also help the high-skilled users to complete high quality portrait sketches different from their customary styles of painting. Note that participants were asked to score their drawing skill using five-point Likert scales.

We have received the participants' comments about system usage, such as, ``I think dualFace is useful, for helping the freehand drawing especially.'' We also received comments about our guidance system, such as, ``Local guidance with mask generation fit my stroke more than the one without it'' and ``Local guidance was surely based on my own, but it looked like a creature.'' All this feedback indicated that mask generation can increase the variation of sketches but sometimes generate an unnatural facial images. This issue can be solved with other neural rendering approaches or a larger face database. We would improve the current prototype to help users draw from different viewpoints and high matching rates with users' drawn strokes.

\subsection{User Satisfaction}
\hzyv{
To verify the user satisfaction whether or not the proposed drawing interface helped users match their objectives, we conducted the user evaluation among three aforementioned interfaces: suggestive drawing UI, shadow drawing UI and ours (Figure 5). We recruited 10 graduate students join this evaluation and a questionnaire was conducted afterwards.  
}

\hzyv{
We confirmed two questions in the questionnaire. The average score for the question ``Do you think your rough sketch is matching with the detail guidance to your expectation with dualFace?" is 4.33 (1, not matched at all; 5, well matched) . The average score for the question ``How would you rate your satisfaction of drawing with dualFace comparing with other two interfaces?" is 4.44. Therefore, dualFace is verified to enable users draw portraits that matches their visions. We also interviewed the participants for further feedbacks on user experiences. The comments on the final drawings include: ``My drawing was better than I though.", ``There were plenty of the details of my drawing which makes it look better.". For usability of dualFace, the users thought that ``It is interesting that it generated the details accordingly." and ``It can automatically generate details, but also beautify the face (drawing).", which are consistent with our findings in Table 1. 
}

\section{Discussion}
\subsection{Computation Cost}

We measured the time cost of portrait drawing for each drawing interface, as shown in Figure~\ref{fig:time}. The minimum time among all sketches using dualFace is 4m15s, and the maximum is 17m15s. Although the users' drawing skill may differ from each other, the drawing results with more time cost lead to better drawing results. The average time cost is around 10min, and the portrait drawings, which cost longer than the average time cost, had more facial details and comparatively better quality than the shorter one. We believe that our local guidance can not only provide enough detailed features for users to follow but also stimulate users' creativity if they intended to spend more time using dualFace.

\subsection{System Interactivity}

\hzy{
Compared with the related sketch to facial image generation approaches
\cite{LinesToFacePhoto19,DeepFacePencil20,EncodinginStyle2020}, the main contribution of dualFace is providing the interactive feedback to users for improving their drawing skills. For these works, users cannot get any help from the systems until the drawing is completed, where users’ essential drawing skills are usually required. Although DeepFaceDrawing can generate high-quality facial image from rough sketch with shadow guidance~\cite{DeepFaceDrawingChenS0XF20}, \hzyv{it is difficult to improve user's drawing skills because they used edge map extracted from images as guidance without separated local-global facial information.} 
In contrast, dualFace can provide interactive sketch support with two-stage guidance for both global features and local facial details. Our system can provide balanced facial information in real-time, so that users can concentrate on learning how to sketch balanced facial contour, especially for novices.
}

\subsection{Generation Diversity}

\begin{figure}
\centering
\includegraphics[width=1.0\linewidth]{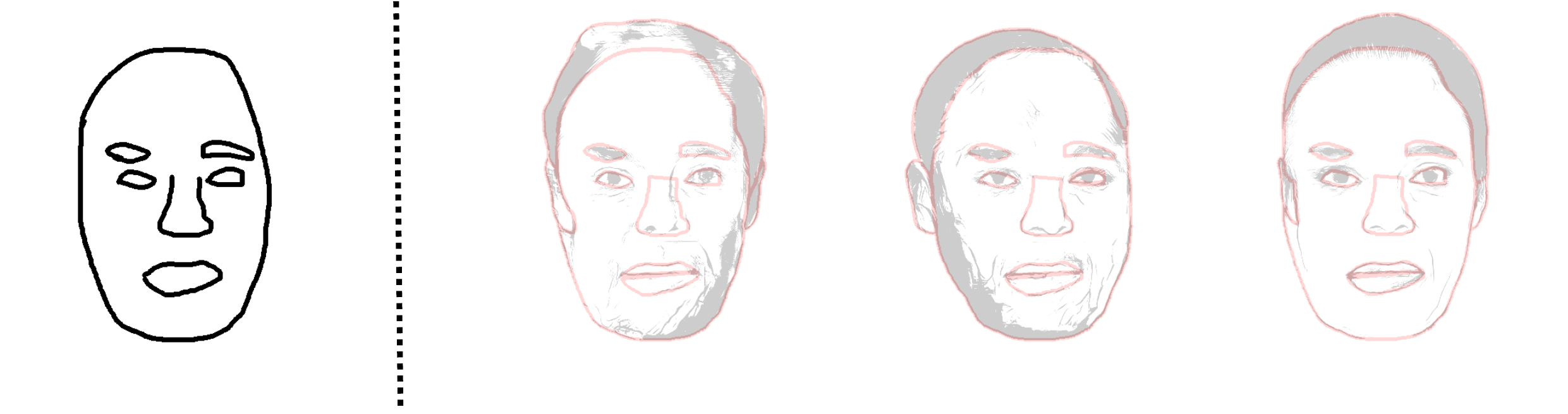}\\
User sketch \quad\quad Local guidance candidates with revised contour

\caption{
Multiple reference candidates (right) generated from the user sketch (left) for local drawing guidance.
}
\label{fig:diversity}
\end{figure}

To meet users’ drawing expectation, it is necessary to ensure generation diversity of facial image database. In this work, the facial diversity could be influenced by database size and mask generation. 
However, the best size of the image database for retrieval on global stage is a hyper-parameter because it is difficult to find out a desirable criterion to evaluate whether generative guidance of dualFace match users' vision automatically for size optimization. 
In our implementation, thus, we selected around 500 typical facial images manually covering different facial types and shapes of facial parts. 
\hzyv{Our selection strategy is to ensure completed facial parts with clear contours in front view and avoid overlapped parts with hair or glasses. 
} 
For mask generation, we can improve the diversity of the generation results for local guidance with multiple references. Figure~\ref{fig:diversity} shows the facial references to users from global stage that users can select the most satisfying generated image as local guidance for facial details drawing. All references maintained the shape restrictions  (red lines) 
from sketch input.

\section{Conclusion}
\label{sec:conclusion}
In this work, we proposed a portrait drawing interface with two-stage global and local guidance. First, we generate a shadow image to provide locations of facial parts when drawing strokes as global guidance. After specifying locations of facial parts as a contour sketch, we then generate detailed facial images from user contour sketches with face mask and portrait drawing generation networks in local guidance. The proposed user interface, dualFace, was verified to be useful \hzyv{and satisfactory} in portrait drawing for users with different levels of drawing skills. We believe that our work contributes to accelerate freehand drawing interfaces.

\renewcommand\wsize{0.195}
\renewcommand\userid{1}
\begin{figure*}
\centering
\begin{minipage}{\wsize\linewidth}\centering

 \subfigure[User Sketch]
 {
\label{fig:imp31}
\includegraphics[width=\textwidth]{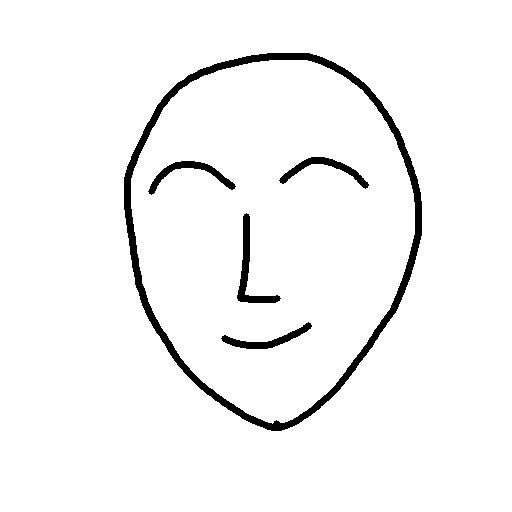}
 }
\end{minipage}
\begin{minipage}{\wsize\linewidth}\centering

 \subfigure[Autocompeleted Mask]
 {
\label{fig:imp32}
\includegraphics[width=\textwidth]{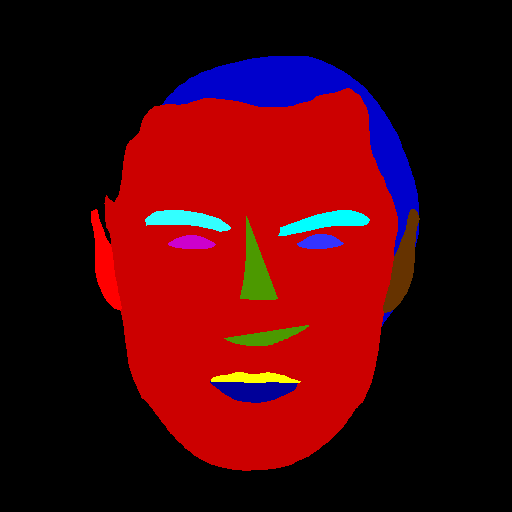}
 }
\end{minipage}
\begin{minipage}{\wsize\linewidth}\centering

 \subfigure[Revised Contour]
 {
\includegraphics[width=\textwidth]{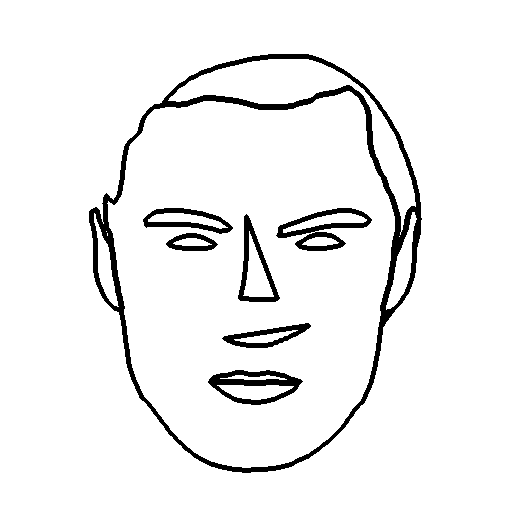}
 }
\end{minipage}
\begin{minipage}{\wsize\linewidth}\centering

 \subfigure[Facial Image]
 {
\label{fig:imp34}
\includegraphics[width=\textwidth]{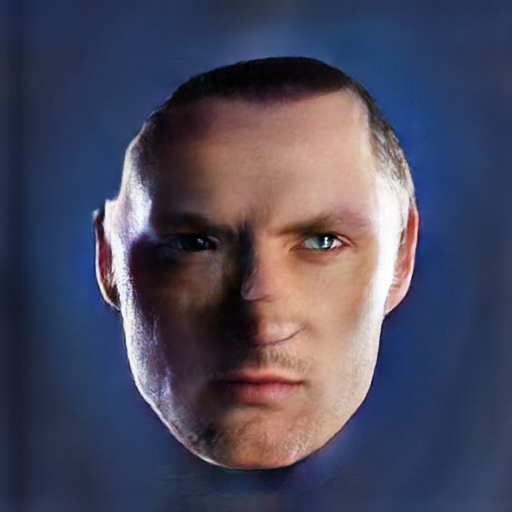}
 }
\end{minipage}
\begin{minipage}{\wsize\linewidth}\centering

 \subfigure[Local Guidance]
 {
\label{fig:imp35}
\includegraphics[width=\textwidth]{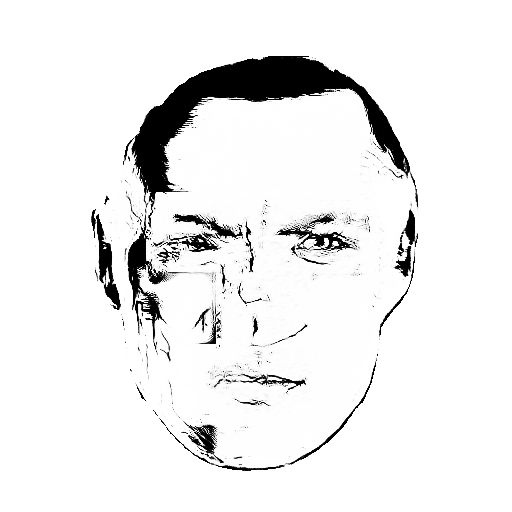}
 }
\end{minipage}
\caption{
Limitations of our work. Abstract sketch may fail to be converted to a reasonable mask (b), where the mouth in user's contour sketch (a) is wrongly regarded as a part of nose. This caused the degeneration of generative image (d) and local guidance (e).
}
\label{fig:imp3}
\end{figure*}

Because the proposed system converts users' sketches to masks by matching the strokes with the example mask, the contour sketch must contain the exact shape information. \hzy{dualFace can only support to draw portrait with realistic style due to real photos in face database.} It is difficult to achieve high-level semantic sketches \hzy{such as emotional faces and exaggerated cartoon style drawing}, because it is challenging to determine the shapes of facial parts currently. If the strokes for facial parts are not closed curves, this may lead to indeterminate contours of facial parts. Figure~\ref{fig:imp3} shows an input sketch \hzy{with a smile face} may generate a strange mask with two separated parts of nose. 
We plan to improve the representation of facial sketches and increase the robustness of dualFace, \hzy{and weigh users' intention and portrait quality.}

\section*{Acknowledgement}
We thank the reviewers for their valuable comments. We also thank Toby Chong for his involvement in idea discussion. This work was supported by Grant from Tateishi Science and Technology Foundation, JSPS KAKENHI grant JP20K19845 and JP19K20316, Japan.

{\small
\bibliographystyle{ieee_fullname}
\bibliography{egbib}
}

\end{document}